\documentclass[traditabstract]{aa}

\usepackage[dvips]{graphicx}
\usepackage{amsmath}

\usepackage{natbib}
\usepackage{txfonts}

\usepackage{amssymb}

\usepackage{subfig}

\newcommand{\comment}[1]{}

\begin{document}

\title{Entropy generation at the multi-fluid MHD solar wind termination shock}
%\titlerunning{The entropy at the multi-fluid solar wind termination shock}

\author{H.-J. Fahr \and M. Siewert}
%\authorrunning{H.-J. Fahr \and M. Siewert}

\institute{Argelander Institut f\"ur Astronomie der Universit\"at Bonn,
Abteilung f. Astrophysik und Extraterrestrische Forschung,
Auf dem Huegel 71, 53121 Bonn (Germany)}

\abstract{In a series of earlier papers, we have developed
expressions for ion and electron velocity distribution functions and their
velocity moments at the passage over the solar wind termination shock. As we
have shown there, with introduction of appropriate particle invariants and
the use of Liouville`s theorem one can get explicit solutions for the
resulting total downstream pressure adding up from partial pressure
contributions of solar wind protons, solar wind electrons and pick-up
protons. These expressions deliver in a first step the main contributions to
the total plasma pressure in the downstream plasma flow and consistently determine
the shock compression ratio. Here now we start
out from these individual fluid pressures downstream of the shock and
thereafter evaluate for the first time the shock-induced entropy production
of the different fluids, when they are passing over the shock
to the downstream side. As is shown here, the resulting ion entropy
production substantially deviates from
earlier calculations using a pseudo-polytropic reaction of the ions to the shock
compression, with polytropies selected to describe fluid-specific
reactions at the shock passage similar to those seen by the VOYAGERs.
From these latter models ion entropy jumps are derived
that depend on the pick-up ion abundance, while our calculations, to the opposite,
deliver an abundance-independent ion entropy production which only depends
on the shock compression ratio and the tilt angle between the upstream
magnetic field and the shock surface normal. We also do show here that only
when including the strongly heated electrons into the entropy balance one
then arrives at the total entropy production that just fulfills the
thermodynamically permitted limit.
}


\maketitle

\section{Introduction}

The plasma physics of shocks in the literature of the past was essentially
reduced to the consideration of flux conservation requirements well known as
Rankine-Hugoniot relations \citep[see e.g.][]{bk-serrinfluid,bk-landaulifshitz,bk-gombosi}
In these relations the internal microphysics of the
shock transition usually is not esplicitly formulated, instead it is
attempted to describe the main shock features with the help of conservation
equations requiring the conservation of the mass - , the momentum- , and the
energy- fluxes at the plasma passage from upstream to downstream side of the
shock. Even though this naturally is the main physical request, this
procedure nevertheless has the normal draw-back that these conservation
relations do not allow for a unique solution, since the fluid-like
conservation requests do not establish a closed system of equations, thus
not allowing for one unique solution. In order to arrive at specific,
discrete solutions one rather has to assume something in addition to the
fluid-like conservation requirements. Often this is done assuming a
polytropic relation between the pressure and density, prescribing for
example the rate of entropy generation at the shock passage.

Things become even much more complicate, if anisotropic plasma pressures and
magnetic field stresses have to be taken into account. Then the system of
conservation equations is substantially enlarged
\citep[see][]{hudson70,bk-baumjohanntreumann,bk-gombosi,erkaev00,bk-diver}
and can only be solved by adding additional informations, such as e.g. two
adiabatic equations requiring the conservation of two CGL- invariants
\citep{cgl56}, as suggested by e.g. \cite{neubauer70-shock-aniso}.
Furthermore, \cite{sf08a,sf09b,fs10b-entropy} have
studied thereafter the action of shock-generated unstable anisotropic
distribution functions that drive magneto-acoustic and Alfv\'{e}nic
turbulences. This was identified as a specific microphysical relaxation
process which effectively operates\ downstream of the shock especially
working in terms of efficient entropy generation.

Furthermore even on the fluid-level the system complicates substantially, if
more than only one plasma fluid have to be considered. If instead of a
monofluid, for instance a multifluid plasma has to be consistently described
at its shock passage, then a number of additional complications have to be
faced in order to arrive at appropriate solutions
\citep[see][]{zankwd93,leroux-fichtner-97-solar-wind-consistent,leroux-fichtner-99-gmir-ts},
with \cite{chalovfahr94,chalovfahr95,chalov-fahr-acr-ts-96} giving
descriptions for two- and three-fluid
plasmas passing over the solar wind termination shock. The three fluids
treated by them as being subject to the bulk motion of the solar wind are
normal solar wind protons (SW`s: eV-energetic), pick-up protons (PUI`s:
KeV-energetic) and anomalous cosmic ray protons (ACRs: MeV-energetic).
A consistent solution of the shock passage of this
three-fluid plasma is only possible, if some additional prescriptions are
made about how these fluids thermodynamically interact with eachother when
undergoing a shock.

In \cite{chalov-fahr-acr-ts-96},
it is formulated that, according to the pickup
proton pressure $P_{pui}$, a specific percentage $\eta$ of these ions is
Fermi-1 accelerated at the shock and injected to the MeV-energetic ACR fluid
regime with an average energy of $E_{inj},$ where both $\eta $ and $E_{inj}$
are unknown parameters which need to be fixed by numbers. This then,
however, has the interesting consequence that PUI`s do not react
adiabatically at the shock, but rather in a quasi-isothermal mode with an effective
polytropic index $\gamma _{pui}$ given by
\begin{equation}
\gamma _{pui}=\gamma _{adia}-\delta (\gamma _{adia}-1)
\end{equation}
where $\gamma _{adia}=5/3$ is the adiabatic index and $\delta $ is given by
\begin{equation}
\delta =\frac{\eta E_{inj}}{(1-\frac{1}{s})P_{pui}\Delta }
\label{eq-delta}
\end{equation}
with $s$ being the compression ratio of the shock.

Thus one can see that the entropy production occuring at the shock passage
in this case is specifically different for the different fluids and is
regulated by specific assumptions for values of $\eta$ and $E_{inj}$ 
\citep[see Fig. 8 of][]{chalov-fahr-acr-ts-96}.

In the following part of the paper we shall now study in detail the entropy
generation in the different fluids when they are passing over the shock,
including electrons as an independent separate fluid. Hereby we try to avoid
the above mentioned ad-hoc assumptions introducing instead kinetic
informations on the single particle behaviour at the shock passage. For that
purpose we first derive the expressions for the downstream pressures of
these separate fluids after taking first a short look into general aspects
of the entropy generation at the shock under Boltzmann kinetic auspices.

\section{The Boltzmann entropic view}

In order to study the entropy of multicomponent systems, we first need
to understand the entropy of a single fluid that is described by an arbitrary
physical velocity distribution function $f(\vec{v})$. Therefore, we need to establish the
basic equations required for this first.

According to e.g. \cite{bk-landaulifshitz}, for a state of a system close to a
thermodynamic equilibrium, assumed to prevail at some distance
upstream and downstream of the shock, the kinetic ensemble entropy per particle can be
given by the expression \citep[see also][]{weizel-58-h-theorem,bk-cercignani,brey-santos-92-entropy,treumann-01-turbulent-plasmas}
\begin{equation}
\bar{S}(\vec{r}) = \frac{S(\vec{r})}{k_B} = -\int f(\vec{r},\vec{v})\cdot \ln (f(\vec{r},\vec{v}))\cdot
d^{3}v,
\label{eq-entropy-boltzmann}
\end{equation}
This equation is valid only in the rest frame of the system, i.e. a reference frame
where the plasma does not possess a bulk flow speed ($\vec{U} = \vec{0}$),
as defined by
\begin{equation}
\vec{U} = <\vec{v}> = \int d^3v \ \vec{v} \ f(\vec{v}).
\end{equation}

For a system in motion, the definition of the entropy has to be modified to account for the
so-called bulk particle flow speed $\vec{U}$,
\begin{equation}
\bar{S}(\vec{r}, \vec{U}) = -\int f(\vec{r},\vec{v}-\vec{U})\cdot \ln (f(\vec{r},\vec{v}-\vec{U}))\cdot
d^{3}v,
\label{eq-entropy-motion}
\end{equation}
since otherwise, and for an arbitrary distribution function $f$, the
definition of the entropy would not be unique. This configuration is
commonly found in systems where the plasma flow is accelerated or decelerated,
and where a ``natural'' reference frame for the system can not be easily found.
One classical example for such a system is an MHD shock wave, where, depending on
the problems under investigation, one could select the rest frame of the
upstream plasma, the downstream plasma or the shock front itself.

\comment{
Another complication that needs to be pointed out is the applicability of
Eqs. \ref{eq-entropy-boltzmann} and \ref{eq-entropy-motion}. In the initial
definition by Boltzmann (1893), the entropy concept was restricted to
systems close to a system close to a thermodynamic equilibrium; today, these
definitions are also commonly applied to arbitrary systems, including those
in a nonthermal equilibrum (``quasi-LTE'') as one commonly finds in systems containing plasma-wave
interactions.
}

In the following parts of the study, we are mainly interested in entropy
jumps at MHD shock waves, where the distribution function $f(\vec{v})$ is
not readily available. Therefore, we need to use a different approach
to the entropy jump. We begin by writing down the differential expression
\begin{equation}
 \Delta \bar{S}
= \frac{\partial \bar{S}}{\partial U}\Delta U
 + \frac{\partial \bar{S}}{\partial T}\Delta T
 + \frac{\partial \bar{S}}{\partial n}\Delta n.
\label{eq-delta-s}
\end{equation}
This expression can be simplified by noting that the kinetic expression
for the Boltzmann entropy is only applicable for quasi-LTE conditions, i.e.
sufficiently relaxed plasma states. Connected with the shock influence
through shock-associated electric and magnetic fields, the plasma properties
may have temporarily attained non-relaxed
intermediate features like asymmetric, anisotropic distributions (which
is the case when using anisotropic MHD jump conditions, see e.g.
\cite{erkaev00,fs06a,sf07a}
or jet-like velocity structures (e.g. due to a possible overshooting of 
electrons or heavy ions, as discussed by \cite{fsc12-multi-shock}).
The standard definition of the Boltzmann entropy can thus evidently not be applied
as long as these intermediate, perturbed, non-relaxed conditions have not yet
reached a quasi-LTE with the help of instabilities driving isotropisation
and relaxation processes. Nevertheless, Eq. \ref{eq-delta-s} may provide a
valuable first-order estimate of the final permitted entropy jump.

Following the line of Boltzmann's understanding, in the above expression the
distribution function should be applied as a normalized one, i.e. as a
velocity-space probability distribution, so that no explicit dependence of
$\Delta \bar{S}$ on the density jump $\Delta n$ appears in the expression for $\bar{S}$.
However, there will be an implicit dependence on $n$ due to the temperature
being directly related to the density, allowing to rewrite Eq. \ref{eq-delta-s}
in the form
\begin{equation}
 \Delta \bar{S}
= \frac{\partial \bar{S}}{\partial U} \Delta U
 + \frac{\partial \bar{S}}{\partial T}\cdot \frac{\partial T}{\partial n}\Delta n.
\label{eq-delta-s-2}
\end{equation}
This result clearly shows the ``kinetic'' (first term on the right) and the
``thermal'' (second term on the right) contributions to the total entropy jump,
while at the same time replacing the temperature jump $\Delta T$ with the density
jump $\Delta n$, allowing us to express the entropy jump as a function
of parameters appearing directly in the jump conditions.

\subsection{The entropy jump for a Maxwell-Boltzmann distribution at the termination shock}

In the following, we make explicit use of two relaxed model distribution functions to describe
the plasma on the upstream and downstream sides of the solar wind termination shock (TS)
in the shock frame. When introducing shock parameters, the subscripts $1$ and $2$ are used to denote
quantities on the upstream and downstream sides of the shock.

First, we assume that the thermal upstream plasma is well described by
a shifted Maxwell-Boltzmann distribution function, i.e.
\begin{equation}
f(\vec{v}) = \frac{C}{T^{3/2}}\exp (-\frac{m(\vec{U}-\vec{v})^{2}}{2 k_{B} T}),
\label{eq-maxwell}
\end{equation}
with a normalisation factor $C=n(m/2\pi k_{B})^{3/2}$. For an MHD shock,
this is just the standard kinetic description that is implicitly assumed to
persist on the upstream side, and - after some relaxation time - also on the
downstream side of the shock.

To simplify the following
calculations, we introduce the short-hand notation
\begin{equation}
\Psi (U,T)=\frac{m(\vec{U}-\vec{v})^{2}}{2k_{B}T},
\end{equation}
which allows us to write down the entropy in the more compact form
\begin{equation}
\begin{split}
\bar{S} &= -\frac{C}{T^{3/2}} \int \exp \left( -\Psi \right)
 \cdot \left[ \ln \frac{C}{T^{3/2}} - \Psi \right] d^3v
\\
&= - \ln \frac{C}{T^{3/2}}
 + \frac{C}{T^{3/2}} \int \exp ( -\Psi) \Psi d^3v.
\end{split}
\label{eq-s-maxwell-0}
\end{equation}
After evaluating the integrals, one simply obtains
\begin{equation}
 \bar{S}(\vec{r})
= \ln (\frac{T^{3/2}}{C})+\frac{1}{2}\Gamma (5/2).
\label{eq-s-maxwell}
\end{equation}
Considering that thermodynamics is only interested in entropy jumps,
but not in absolute values, the constant second term in this sum can
be interpreted as a normalisation constant \citep[see e.g.][]{collier-95-kappa-transport}.

This expression in principle allows us to derive the entropy jump
$\Delta \bar{S} = \bar{S}_2 - \bar{S}_1$ between both sides of the shock, assuming that
we have solved the MHD jump conditions.
However, it is also possible to derive a more explicit expression for the
entropy jump that does not depend on explicit values for the shock parameters. Using
Eq. \ref{eq-s-maxwell-0}, it is possible to evaluate the partial derivatives
in Eq. \ref{eq-delta-s-2}. After some elementary operations, we obtain
the relations
\begin{equation}
 \frac{\partial }{\partial T}\Psi(U,T)
= -\frac{1}{T}\Psi (U,T)
\end{equation}
and
\begin{equation}
 \frac{\partial }{\partial U}\Psi (U,T)
= 2 \sqrt{\frac{m}{2 k_B T}} (U - v\cos\theta),
\end{equation}
where $\theta = \angle(\vec{U},\vec{v})$. Using these relations, we obtain
\begin{equation}
 \frac{\partial }{\partial T} \bar{S}
= \frac{3}{2T} 
 + \int \frac{\partial}{\partial T} \frac{C}{T^{3/2}}\exp (-\Psi (U,T))\Psi (U,T)d^{3}v.
\end{equation}
Evaluating the partial derivatives and collecting terms, we further obtain
\begin{equation}
\begin{split}
 \frac{\partial }{\partial T} \bar{S}
=& \frac{3}{2T}
\\
 &- \frac{5}{2T} \int \frac{C}{T^{3/2}}\exp (-\Psi (U,T))\Psi (U,T)d^{3}v
\\
 &+ \frac{1}{T} \int \frac{C}{T^{3/2}}\exp (-\Psi (U,T))\Psi^2(U,T) d^{3}v.
\end{split}
\end{equation}
After some elementary substitutions, the integrals evaluate to
\begin{equation}
 \frac{\partial }{\partial T} \bar{S}
= \frac{1}{T} \cdot
 \left(\frac{3}{2} - \frac{5}{4} \Gamma\left(\frac{5}{2}\right) + \frac{1}{2}\Gamma\left(\frac{7}{2}\right)\right)
= \frac{3}{2T}.
\end{equation}
Using a similar approach, we obtain for the other partial derivative
\begin{equation}
\begin{split}
 &\frac{\partial }{\partial U} \bar{S}(\vec{r})=
\\
 & -2\sqrt{\frac{m}{2 k_B T}} \int \frac{C}{T^{3/2}} (U-v\cos\theta) \exp(-\Psi (U,T))\Psi (U,T)d^{3}v
\\
 &+ 2 \sqrt{\frac{m}{2 k_B T}} \int \frac{C}{T^{3/2}} (U-v\cos\theta) \exp(-\Psi(U,T))d^{3}v.
\end{split}
\end{equation}
After introducing spherical coordinates and substituting $w = U - v\cos\theta$, this
expression evaluates into
\begin{equation}
 \frac{\partial }{\partial U} \bar{S}
= - \sqrt{\frac{m}{2 k_B T}} (\Gamma(3) - \Gamma(2))
= - \sqrt{\frac{m}{2 k_B T}}.
\end{equation}

Therefore, we obtain an entropy jump for the general Maxwell-Boltzmann distribution
given by
\begin{equation}
 \Delta \bar{S}
= - \sqrt{\frac{m}{2 k_B T}} \Delta U
  + \frac{3}{2T} \frac{\partial T}{\partial n} \Delta n.
\label{eq-delta-s-mb}
\end{equation}
The only unknown parameter entering this expression is the thermodynamic
relation between temperature and density, which reflects the entire
microphysics inside the shock transition layer that is inaccessible to MHD.

Depending on the behaviour of $\frac{\partial T}{\partial n}$, the entropy
jump may be negative, suggesting that the shock transition is physically
impossible. However, one also has to consider that the partial derivative
essentially represents an average of the entire microphysics of the shock
transition, to which the single fluid approximation just may be a too
strong simplification, and the presence of plasma-wave interactions or
heavy ions in principle even allow a decrease in the entropy in a single
part of the entire system (see also Eq. \ref{eq-delta}).

In the limit of a single component shock, one can request
the entropy change to be positive ($\Delta \bar{S} > 0$), and obtains
\begin{equation}
\frac{\partial T}{\partial n} > \frac{2}{3} \sqrt{\frac{m}{2 k_B T}} T \frac{\Delta U}{\Delta n},
\end{equation}
which can be simplified further by introducing the
MHD compression ratio $s=\frac{U_1}{U_2} = \frac{n_2}{n_1}$, resulting in
\begin{equation}
\begin{split}
\frac{\partial T}{\partial n}
 &> \frac{2}{3} \sqrt{\frac{m}{2 k_B T}} T U_1 \frac{s^{-1}-1}{s-1}
\\
 &= \frac{2}{3} \sqrt{\frac{m T}{2 k_B}} U_2 \frac{1-s}{s-1}.
\\
 &= -\frac{2}{3} \sqrt{\frac{m T}{2 k_B}} U_2.
\end{split}
\label{eq-entropy-cond}
\end{equation}
This equation provides an easy approach to the question whether a
physical, entropy-increasing shock is possible without having to introduce
additional nontrivial thermodynamic degrees of freedom to the shock transition.

\subsection{The entropy of a $\kappa$ function}

The shock passage very likely provokes nonthermal equilibrum conditions,
and therefore, a distribution function $f(\vec{v})$ that differs from the
classical thermodynamic Maxwellian (Eq. \ref{eq-maxwell}) is required.
Following theoretical arguments studied by e.g. \cite{treumann99-scripta,treumann-jaroschek-scholer-04-plasma},
or \cite{livadiotis-12-non-equilibrum}
we adopt a nonthermal $\kappa$ function to describe
the quasi-stable nonthermal downstream equilibrum state that likely develops
on the near downstream side of the shock. Therefore, we need to understand the entropy
stored in a $\kappa$-function, which allows us to compare it to the entropy
gain using a conventional Maxwellian distribution (Eq. \ref{eq-delta-s-mb}).

Taking the standard definition of an isotropic $\kappa$ distribution,
\begin{equation}
 f_\kappa(\vec{v})
= \frac{n}{(\pi \sqrt{\kappa} \Theta^{2})^{3/2}}
 \frac{\Gamma(\kappa+1)}{\Gamma (\kappa-3/2)}\left[1+\frac{v^{2}}{\kappa\Theta^{2}}\right]^{-(\kappa+1)}.
\end{equation}
with
\begin{equation}
\Theta^2 = \frac{2 kT_c}{m},
\end{equation}
we are able to calculate the $\kappa$-entropy using the same Boltzmann formalism
as used above for the Maxwell-Boltzmann distribution,
as long as one assumes that the $\kappa$-function does reflect an equilibrum state between
classical elastic scattering and energy diffucion \citep[see e.g.][]{collier-95-kappa-transport}.
Under these conditions, the entropy of a $\kappa$ distribution function
is given by
\begin{equation}
\begin{split}
 \bar{S}^{\kappa }
=& \ln \{\frac{(\Theta ^{2}\pi \kappa )^{3/2}\Gamma (\kappa
-1/2)}{\Gamma (\kappa +1)}
\\
&\cdot \exp [(\kappa +1)(\digamma (\kappa +1)-\digamma
(\kappa -1/2))]\},
\end{split}
\end{equation}
where the digamma function $\digamma $ is defined by:
\begin{equation}
\digamma (z)=\frac{d}{dz}\ln (\Gamma (z)).
\end{equation}
Using elementary properties of the gamma function
(see e.g. Abramowitz \& Stegun), the somewhat unhandy expression can
be simplified by first eliminating the $\digamma$ symbols, leading to
\begin{equation}
\begin{split}
 \bar{S}^{\kappa }
=& \ln \{\frac{(\Theta ^{2}\pi \kappa )^{3/2}\Gamma (\kappa-1/2)}{\Gamma (\kappa +1)}
\\
 &\cdot \exp [(\kappa +1)(\frac{\kappa \Gamma (\kappa )}{\Gamma (\kappa +1)}-\frac{(\kappa -3/2)\Gamma (\kappa -3/2)}{\Gamma (\kappa-1/2)})]\}.
\end{split}
\end{equation}
After some further evaluations, the $\gamma$ functions in the exponent
cancel out, leaving only
\begin{equation}
 S^{\kappa }/k_{B}
= \ln \{\frac{(\Theta ^{2}\pi \kappa )^{3/2}\Gamma (\kappa-1/2)}{\Gamma (\kappa +1)}\}.
\label{eq-entropy-kappa}
\end{equation}
One specifically interesting property of this expression is the limit for
$\kappa \rightarrow \infty$, which should reproduce the classical Maxwellian limit
(Eq. \ref{eq-s-maxwell}). Using Stirlings formula (see e.g. Abramowitz \& Stegun),
one easily sees that
\begin{equation}
\frac{\Gamma (\kappa-1/2)}{\Gamma (\kappa +1)}
 \stackrel{\kappa \rightarrow \infty}{\rightarrow}
 \kappa^{-3/2},
\end{equation}
and the entropy becomes
\begin{equation}
 \bar{S}^{\kappa }
\rightarrow \ln (\Theta ^{2}\pi)^{3/2},
\end{equation}
which (with the exception of a different normalisation, see \cite{collier-95-kappa-transport}) is just
the entropy for a Maxwellian distribution function.

\subsection{The classical MHD shock entropy gain and the polytropic index $\gamma$}

For completeness, we also briefly mention the entropy gain found
in most MHD textbooks, as is derived from classical shock relations,
and without paying much attention to the kinetic origin of the entropy
\citep[see e.g.][]{bk-serrinfluid,bk-landaulifshitz}.
This jump in entropy per unit mass at the shock, judged in
the frame of the bulk plasma flow, is given by
\begin{equation}
 \Delta \bar{S}_{MHD}
= \ln [\frac{P_{2}}{P_{1}}(\frac{\rho _{1}}{\rho _{2}})^{\gamma_{adia}}],
\label{eq-delta-s-mhd}
\end{equation}
where $\gamma_{adia} = 5/3$ is the so-called adiabatic index.
For this specific choice of $\gamma_{adia}$,
it immediately follows that a gas, reacting
strictly adiabatically at the shock compression, will not increase
its thermal entropy in the bulk frame at all, which strongly suggests that
the shock transition can not be purely adiabatic.

However, the adiabatic index $\gamma_{adia}$ is just a special case
of the more general polytropic relation, which for the system studied
here can be represented in the form
\begin{equation}
T = T_{0}\cdot (n/n_{0})^{\gamma -1},
\end{equation}
where $\gamma$ is the more general polytropic index, for which $\gamma = \gamma_{adia}$
is just one special value. In addition to the obvious impact on the MHD
entropy jump (Eq. \ref{eq-delta-s-mhd}), this equation also allows
to quantify the previously unknown parameter in Eq. \ref{eq-delta-s-mb},
i.e. the partial derivative $\frac{\partial T}{\partial n}$, which for
the general polytropic relation becomes
\begin{equation}
 \frac{\partial T}{\partial n}
= (\gamma -1)(\frac{n}{n_{0}})^{\gamma -2}\frac{T_{0}}{n_{0}}.
\label{eq-polytropic-d}
\end{equation}
Observational data studies usually treat the polytropic index $\gamma$
as a free fit parameter, while theoretical studies often apply this relation
as an ``ad-hoc'' boundary condition; in principle, for arbitrary shocked systems,
the $T$-$n$ relation could be of a form that differs from this model approach.
However, at an MHD shockwave, this approach can be justified by the fact that
the microphysics of the shock can not be modeled by a fluid theory, and therefore,
a polytropic index can be interpreted as an averaged description of the microphysical
plasma-wave interactions in the shock transition layer.
Applying Eq. \ref{eq-polytropic-d} to Eq. \ref{eq-entropy-cond}, one easily sees that
the monofluid polytropic approach to the MHD shock always increases the
entropy in the system for $\gamma > 1$, which covers pretty much all polytropic
indices found in the literature (where, usually, $\gamma = 1... 2$).

In this study, we present one possible way of connecting an effective
polytropic index with a multifluid MHD shock wave, thus introducing
theoretical concepts to the ad-hoc boundary condition,
and compare the result with the best fitting effective polytropic index found
by \cite{wu-09-ion-heating} at the solar wind termination shock.

\section{The multi-fluid plasma at the termination shock}

\subsection{Pressures in a multifluid system}

As suggested by Eq. \ref{eq-delta-s-mhd}, the upstream and downstream pressures
of the shocked plasma provide an important quantity for the entropy problem.
For an MHD approach, the upstream and downstream pressures directly enter
Eq. \ref{eq-delta-s-mhd}, while for the kinetic approach (Eq. \ref{eq-delta-s-2}),
we need to convert between pressures (that appear in the jump conditions)
and temperatures (that appear in the kinetic distribution function).
Due to the fact that the immediate downstream side of the shock may
represent a thermal nonequilibrum, this conversion can be difficult;
in this paper, we assume that the downstream side
is defined as the region where a new (possibly nonthermal) equilibrum
has been reached, so that we can ignore more details of the shock transition.
However, we do not ignore the multifluid characteristics of the shock,
for which we adopt an overshooting description, i.e. a description
where particles possessing different electric charges and masses
will react differently to the global electric ramp of the shock potential.
While thermal and PUI protons do not require this additional detail,
it becomes important when we study the entropy gain of the electron component
in Sect. \ref{sec-electrons}.

In this study, we consider MHD shocks in a one-dimensional approach with the
shock normal $\vec{n}$ assumed to be parallel to the upstream bulk flow
velocity $\vec{U}_{1}$ and an upstream magnetic field vector tilted by an
angle $\alpha $ with respect to $\vec{n}$. We consider three different
fluids, namely solar wind protons, solar wind electrons and pick-up ions.
For an explicit description of the downstream pressure $P_{2,p}$ of the
thermal SW proton component, we adopt the relations given by
\cite{fsc12-multi-shock} and \cite{fs13-multi-p},
\begin{equation}
P_{2,p}=\frac{1}{3}s\cdot (2A(s,\alpha )+B(s,\alpha ))\cdot P_{1,p},
\label{eq-p2p-prot}
\end{equation}
where the indices 1, 2 denote upstream and downstream quantities,
respectively, $s=U_{1}/U_{2}$ denotes the shock compression ratio, and the
functions $A(s,\alpha )$ and $B(s,\alpha )$ are given by
\begin{equation}
A(\alpha )=\sqrt{\cos ^{2}\alpha +s^{2}\sin ^{2}\alpha}
\end{equation}
and
\begin{equation}
B(\alpha )=s^{2}/A^{2}(\alpha ),
\end{equation}
where the angle $\alpha$ defines the inclination between the shock surface
normal $\vec{n}$ and the upstream magnetic field $\vec{B}_{1}$ (i.e. for a
perpendicular shock this means $\alpha =\pi /2)$.

In our multifluid description, the downstream PUI pressure $P_{2,pui}$ is derived
in the same way as the downstream proton pressure $P_{2,p}$, i.e. the main
difference between both pressures is a factor $\Gamma_{1}=P_{1,pui}/P_{1,p}$,
and therefore is simply given by the following analogous formula
\begin{equation}
\begin{split}
 P_{2,pui}
&= \frac{1}{3}s[2A(\alpha )+\frac{s^{2}}{A^{2}(\alpha )}]\cdot P_{1,pui}
\\
&=  \frac{1}{3}s[2A(\alpha )+\frac{s^{2}}{A^{2}(\alpha )}]\cdot \Gamma_1 P_{1,p}.
\end{split}
\label{eq-p2p-pui}
\end{equation}
This approach turns out to be justified because a pick-up proton cannot be physically
differentiated from a solar wind proton as a different ion species, due to the
following reasons.
The main complication when dealing with an initial distribution of PUIs at a shock
is the reflection of a certain fraction of the energetic ions from the electric
shock potential. Considering the total velocity $\vec{u}$ of an individual ion in the
shock frame, i.e. $\vec{u} =\vec{v} + \vec{U}$, it becomes obvious that
some ions may possess a velocity vector that does not enable them to cross
the shock potential at the first attempt. Instead, these ions are
reflected into the shock precursor region where they induce
local two-stream instabilities. Following this, they do gain energy and momentum
from interactions with just these instabilities, until they finally get transported
across the shock. In many MHD shock simulations, a different (i.e. purely numerical)
approach to this situation
has been studied, but the answer concerning the resulting final
downstream plasma mixture has not yet been conclusively given
\citep[see][]{scholer-93-shock-par,liewer-et-al-93-t-shock,kucharek-scholer-95-injection,kucharek-06-ts,zank10-ts,matsu11-ts-simulation,wu10-hybrid-simulations}.
Results obtained in these simulations strongly
depend on the shock compression ratio and especially on the upstream PUI
velocity distribution used by the authors at the start of their simulations, e.g.
cooled or heated shell distributions. Nevertheless, these simulations clearly
demonstrate that a coupling between PUIs and thermal ions is introduced at the shock in
a natural way.

To get a reliable answer to this problem for our purposes here we look back
at the work by \cite{chalov-fahr-acr-ts-96}, who studied the kinetic transport
of a statistical sample of PUIs at their passage over the
electric and magnetic shock structure, starting from a realistic
upstream PUI distribution functions taking into account the cooling
of PUIs before they enter the shock
according to most up-to-date theories \citep[see][]{chalovfahr95}.
Following the method by \cite{decker88}
involving a de Hoffmann-Teller
frame and the conservation of the magnetic moment, they found that
\citep[][Figr. 6 and 7]{chalov-fahr-acr-ts-96}, for realistic upstream PUI
distributions, the fraction of reflected (second order PUIs) over directly
transmitted (first order PUIs) is less than $10^{-2}$ and is especially low
for perpendicular shocks ($\alpha =\pi /2)$. Taking this result as solid, it
is thus possible to assume for the rest of this paper that PUIs behave practically
the same way as SW ions, and that they are all transmitted through the shock,
with just a negligibly small number of reflected PUIs not taken into account
in our present consideration. We study the entropy gain in this approximation
of the multicomponent shock in Sect. \ref{sec-magnet}.

\subsection{The joint downstream ion distribution}

\label{sec-joint}

In addition to this straightforward approach, we will also apply a different
description of the downstream plasma, motivated by the coupling between thermal
and nonthermal ions that is naturally introduced at the shock.
Since both downstream ion populations (i.e. thermal protons and PUIs)
are located at overlapping regions in phasespace, it is possible to
describe them as one joint downstream pick-up ion and solar wind proton
distribution. As demonstrated by \cite{fs13-multi-p},
the main features of this combined ion distribution can
be represented surprising well by a joint Kappa distribution,
\begin{equation}
f_{2}(v)=\frac{n_{2}}{(\pi \kappa _{2}\Theta _{2}^{2})^{3/2}}\frac{\Gamma
(\kappa _{2}+1)}{\Gamma (\kappa _{2}-3/2)}[1+\frac{v^{2}}{\kappa _{2}\Theta
_{2}^{2}}]^{-(\kappa _{2}+1)},
\end{equation}
with a Gaussian core velocity spread $\Theta _{2}$ and a net Kappa index
$\kappa _{2}$ as characteristic parameters. We interprete the downstream
solar wind proton population as constituting the so-called Gaussian core
$\Theta _{2}$ of the Kappa distribution
\citep[following][]{collier-95-kappa-transport,heerikhuisen08-kappa,liv-mcc-09-beyond-kappa}
and thus fix the
needed Kappa function parameters as done by \cite{fs13-multi-p}. The
downstream thermal width $\Theta _{2}$ of the Gaussian core then turns out as
\begin{equation}
\Theta _{2}^{2}=\frac{s}{3}(2A(\alpha )+B(\alpha ))\frac{2}{n_{2}m}P_{1,p}.
\end{equation}
The joint Kappa index $\kappa _{2}$ follows from the requirement of
the pressure identity $P_{2}^{\kappa }=P_{2,p}+P_{2,pui}$, resulting in
\begin{equation}
\kappa _{2}=\frac{3[1+\zeta K]}{2\zeta \lbrack K-1]}
\end{equation}
with the parameter $K$ given by
\begin{equation}
K=\frac{\kappa _{1,p}-3/2}{\kappa _{1,pui}-3/2}\frac{\kappa _{1,pui}\Theta
_{1,pui}^{2}}{\kappa _{1,p}\theta _{1,p}^{2}},
\end{equation}
where the upstream PUI abundance is given by $\zeta = n_{1,pui}/n_{1,p}$.
In this description, we assume separate Kappa distribution functions on the upstream
side of the TS, one for the thermal ions (using the parameters $\Theta_{1,p}$
and $\kappa_{1,p}$), and another separate function for the PUIs (using
the parameters $\Theta_{1,pui}$ and $\kappa_{1,pui}$). This allows for a greater
flexibility in upstream configurations, with $\kappa \rightarrow \infty$ for a
pure Maxwellian distribution (i.e. the thermal component),
and $\kappa \rightarrow 3/2$ for a pure $v^{-5}$ power law (i.e. the PUI component).
In the following, we will adopt the same parameters as \cite{fs13-multi-p},
resulting in the parameter $K = 119$.

We study the entropy gain at the multicomponent shock in this approach
in greater detail in Sect. \ref{sec-kappa}.

\subsection{Downstream pressures in magneto-adiabatic or pseudo-polytropic approaches}

Before calculating actual entropy gains, we study the alternate
approach by \cite{wu-09-ion-heating}, where the relation between downstream and upstream PUI
pressures is obtained as a pseudo-adiabatic reaction of the PUIs to the
shock compression given by
\begin{equation}
 P_{2,pui}
= (\rho _{pui,2}/\rho _{pui,1})^{\gamma} P_{1,pui}
= (s)^{\gamma_{p}}P_{1,pui},
\end{equation}
where instead of a joint $\kappa$ index or a kinetically derived
enhancement factor, a PUI-specific \textit{p}olytropic
index $\gamma_p \geq \gamma_{adia} = 5/3$ is used to describe an
additional heating of PUIs at the shock passage, which in the view
of the authors is a need to make the simulation results
better fit the Voyager-2 shock data \citep[see][]{richardson08-voy2-accel}.
We can now compare their approach with our kinetic model by deriving
the adequate polytropic
index $\gamma_{p}$ that would give equivalent PUI pressure
transformations as derived from our model. This requirement leads to
the following expression:
\begin{equation}
 \Pi (\alpha ,s)
= \frac{1}{3}s[2A(\alpha )+\frac{s^{2}}{A^{2}(\alpha )}]
= (\rho_{pui,2}/\rho _{pui,1})^{\gamma_{p}}
= s^{\gamma_{p}},
\label{eq-pi-gamma}
\end{equation}
or, for the resulting polytropic index $\gamma_{}$,
\begin{equation}
\gamma_{p} = \frac{\ln(s(\frac{2}{3} A(s,\alpha) + \frac{s^2}{3 A^2(s,\alpha)}))}{\ln s}.
\label{eq-final-gamma}
\end{equation}

\begin{figure}
\includegraphics[width=\columnwidth]{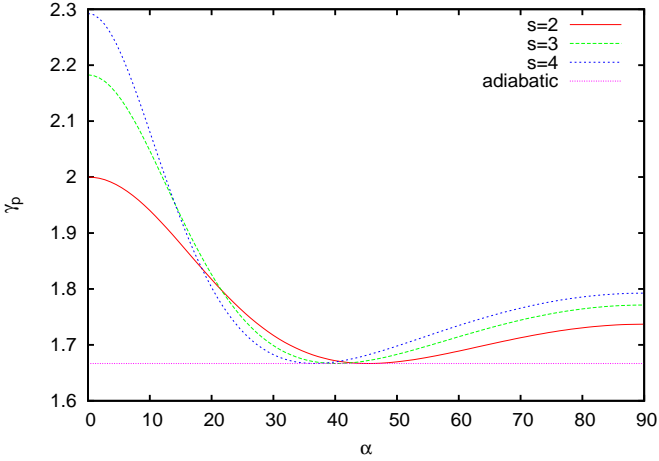}
\caption{Polytropic indices obtained from our kinetic shock model according to Eq. \ref{eq-final-gamma}
for different magnetic field orientations and MHD compression ratios.}
\label{fig-gamma}
\end{figure}

For a quasiperpendicular shock as encountered by the Voyager-2 spacecraft,
we need to adopt $\alpha =90^\circ$, where
\begin{equation}
A(s,\alpha) \rightarrow s,
\end{equation}
and Eq. \ref{eq-final-gamma} reduces to
\begin{equation}
\gamma_{p} = \frac{\ln(s(\frac{2s}{3}+\frac{1}{3}))}{\ln s}.
\end{equation}
For a compression ratio of $s=3$, we simply obtain
\begin{equation}
 \gamma_{p}(90^\circ)
= \frac{\ln (7)}{\ln (3)} = 1.77.
\end{equation}
Evaluating the same equation for a more parallel shock, e.g. with $\alpha =20^\circ$,
we instead obtain $\gamma_{p}(20^\circ) = 1.82$. Global results for all
angles and various compression ratios are presented in Fig. \ref{fig-gamma},
which demonstrates that, for most magnetic field orientations, our
kinetic model for the shock transition results in an over-adiabatic behaviour.
As demonstratd by Figure \ref{fig-gamma}, this overadiabatic behaviour with $\gamma_{p} > \gamma_{adia}=5/3$,
dominates the parameter region for angles $\alpha < 30^\circ$ and $\alpha > 40^\circ$. However, for angles between $30^\circ < \alpha < 40^\circ$, the effective polytropic indices are unexpectedly close to the classical adiabatic value of $\gamma_{p} \simeq \gamma_{adia)}$. The reason for this is not directly evident from the calculations presented here, but can be understood with the help of the results published earlier by \cite{fs10b-entropy}. This earlier paper demonstrates that the range of tilt angles between $30^\circ$ and $40^\circ$ is characterized by the absence of downstream ion temperature anisotropies (i.e. $T_\perp \simeq T_\parallel$),
which can be easily seen on Fig. 3 in the mentioned earlier study. This means that, on this narrow intervall, the two degrees of freedom parallel and perpendicular to $\vec{B}$ are equally heated, which is the same behaviour as the one found in the case of an unmagnetized gas. In addition, the changes of the temperature from upwind to downwind behave just as in the adiabatic shock compression, i.e. $P_2/P_1 = (n_2/n_1)^{\gamma_{adia}}$.

These values display the same behaviour found by \cite{wu-09-ion-heating},
who found a general over-adiabatic behaviour
when trying to best-fit the Voyager-2 shock observations.
They obviously needed a preferential heating of the
PUI-fluid compared to the SW fluid, and in their case obtain it by a
fluid-specifically increased polytropic index.

While this may suggest that the two theoretical
approaches can both deliver similar results, it must, however, be recognized
that the selected $\gamma_{p}$-value invokes an unexplained ad hoc
process for PUIs. This follows from the fact that this approach
treats the PUI protons and their thermodynamic reaction to a shock
compression in a substantially different way
from that of the normal solar wind protons. The justification for this approach may be
that some of the PUI protons are reflected by the shock and later get
transmitted after experiencing some energy gain. On the other hand, seen
from physical grounds and argued on the basis of results presented by \cite{chalov-fahr-acr-ts-96},
who found that PUI reflection is fairly unlikely, protons
should react like protons, disregarded whether they are of the PUI or of the
solar-wind type. In our approach, this results in PUIs being heated more
efficient than solar wind protons due to the simple fact that PUIs are
already hotter upstream of the shock. This means that protons of both fluids
in fact do react completely alike, and the resulting pressure conversion
simply is derived under conservation of kinetic particle invariants. This
difference in the shock reaction also leads to different entropy production
rates, as we shall demonstrate in the remaining sections.
Especially the entropy generated in the pseudo-polytropic multi-ion shock
turns out to be very much different from corresponding results that we
obtain when using our ``magneto-adiabatic'' approach.
We present explicit values for the entropy generated in a
pseudo-polytropic multi-ion shock in Sect. \ref{sec-pseudo}.

\section{The entropy production at the multifluid shock}

\subsection{The entropy jump in the ``magneto-adiabatic'' approach}

\label{sec-magnet}

\begin{figure}
\includegraphics[width=\columnwidth]{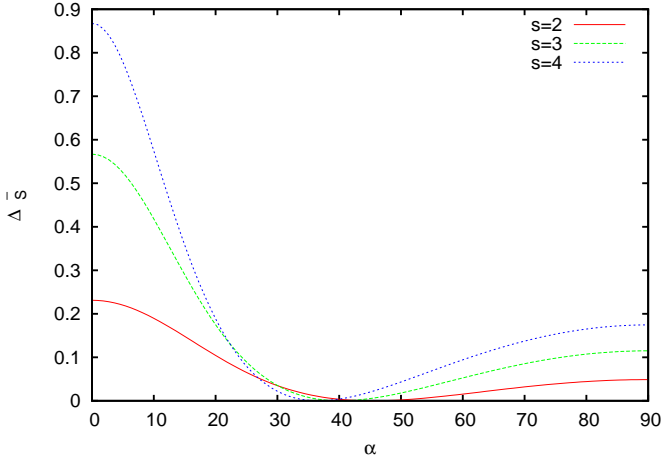}
\caption{The normalized ion entropy gain as a function of the magnetic field tilt
angle $\alpha$ and the compression ratio $s$ for the magnetio-adiabatic approach.}
\label{fig-deltas-magnet}
\end{figure}

Using our ``magneto-adiabatic'' formulae for the downstream ion pressures
(Eq. \ref{eq-p2p-prot}), we are now able to calculate the following non-vanishing entropy jump
in the MHD limit (Eq.\ref{eq-delta-s-mhd}):
\begin{equation}
 \Delta \bar{S}
= \ln [\frac{s}{3(1+\zeta K)}(2A+B)(1+\zeta K)(\frac{\rho _{1}}{\rho _{2}})^{\gamma_{adia}}],
\label{eq-deltas-magnet-0}
\end{equation}
where $\zeta = n_{pui}/n_{p}$ is the PUI abundance, and $K$ is a parameter reflecting
the upstream thermal and nonthermal proton configuration (see Sect. \ref{sec-joint}
for a more detailed explanation). Interestingly enough, this
equation simplifies to an expression independent on
the PUI abundance (and the upstream parameter $K$), simply given by: 
\begin{equation}
 \Delta \bar{S}
= \ln [\frac{s}{3}(2A+B)s^{-\gamma_{adia}}]
\label{eq-deltas-magnet}
\end{equation}
This expresses the expected fact that the entropy production in our case
does not depend on the abundance $\zeta$ of upstream pick-up ions,
which trivially follows from the concept that PUI protons at the TS
should behave exactly like solar wind protons, i.e. depending only on
the compression ratio $s$ and the magnetic tilt angle $\alpha$.
A graphical representation of the entropy gain for this pressure model
is given in Fig. \ref{fig-deltas-magnet}.

We can also compare this result with the effective polytropic indices $\gamma_p$
presented on Fig. \ref{fig-gamma}. Adopting a description using polytropic
indices, we obtain
\begin{equation}
 \Delta \bar{S}
= \ln [s^{\gamma_p-\gamma_{adia}}].
\end{equation}
This relation easily proves that, for $\gamma_p < 8/3$, the normalized entropy increase
is of the order of $\ln s$, i.e. between 0 and 1.38, which agrees with the numerical results
given in Fig. \ref{fig-deltas-magnet}.

\subsection{The entropy jump for non-adiabatic PUIs}

\label{sec-pseudo}

\begin{figure}
\includegraphics[width=\columnwidth]{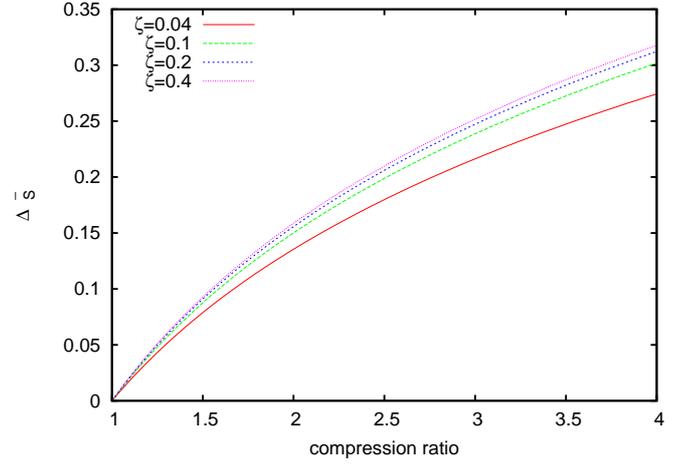}
\caption{The normalized ion entropy gain as a function of the PUI abundance $\zeta$
and the PUI polytropic index $\gamma$ for a shock where
thermal protons behave adiabatically, and the PUIs behave non-adiabatically.
The compression ratio is $s=3$.}
\label{fig-deltas-wu}
\end{figure}

In the description by \cite{wu-09-ion-heating}, however, the entropy jump explicitly depends
on the pick-up ion abundance $\zeta$ as we will demonstrate now. Following
these authors, solar wind protons and pick-up ions do react to the shock
compression in different polytropic forms, the first characterized by a
polytropic index $\gamma_{p}$, the latter by a larger pseudo-polytropic
index $\gamma _{pui}$. Thus, when looking for the related proton entropy jump
of the joint ion population, one finds the following result that is valid for
these multi-polytropic conditions:
\begin{equation}
 \Delta \bar{S}^{p}
= \ln [\frac{P_{p1}s^{\gamma _{p}}+P_{pui,1}s^{\gamma _{pui}}%
}{P_{p1}+P_{pui,1}}(s^{-\gamma _{adia}})]
\end{equation}
Now, introducing the same representation of the upstream pressures $P_{pui,1}$
and $P_{p,1},$ as used in our approach, and again introducing the PUI abundance
$\zeta$, we obtain an entropy jump given by
\begin{equation}
 \Delta \bar{S}^{p}
= \ln [\frac{s^{\gamma _{p}}+\zeta Ks^{\gamma _{pui}}}{1+\zeta K}
 s^{-\gamma _{adia}}].
\end{equation}
We now assume that solar wind protons are reacting adiabatically at the shock,
so we can select $\gamma_p = \gamma_{adia}$, and obtain
an entropy jump of
\begin{equation}
 \Delta S^{p}
= k_{B}\ln [\frac{1+\zeta Ks^{\gamma _{p}-\gamma _{adia}}}{1+\zeta K}],
\label{eq-deltas-wu}
\end{equation}
making it evident that their expression inherently depends on
the PUI abundance $\zeta$.

A graphical representation of the entropy gain in this model is presented in
Fig. \ref{fig-deltas-wu}, where one easily sees that $\Delta \bar{S}^W$ is about
one order of magnitude smaller than the magneto-adiabatic entropy gains presented in
the previous section. This can be understood easily, as the entropy
gain in the non-adiabatic PUI desciption is exclusively related to the
nonadiabatic PUI behaviour, which only make a fraction of the entire
entropy of the system. Thermal ions, behaving adiabatically, do not increase
their entropy at all in this representation.

The selection of this assumption was based on a best fit approach to the Voyager data, which
does not allow to assess the behaviour of the thermal solar wind plasma component.
Therefore, any theoretical modeling must take great care when making model
assumptions concerning the behaviour of the thermal protons. However,
our results suggest that, the thermal protons do most likely not behave
adiabatically at the shock (unless average tilt angles of $\alpha \simeq 40^\circ$ are
assumed, see Fig. \ref{fig-gamma}), as this behaviour obviously results in a strong suppression
of entropy production.

\subsection{The entropy jump for a multifluid system including electrons}

\label{sec-electrons}

\begin{figure}
\includegraphics[width=\columnwidth]{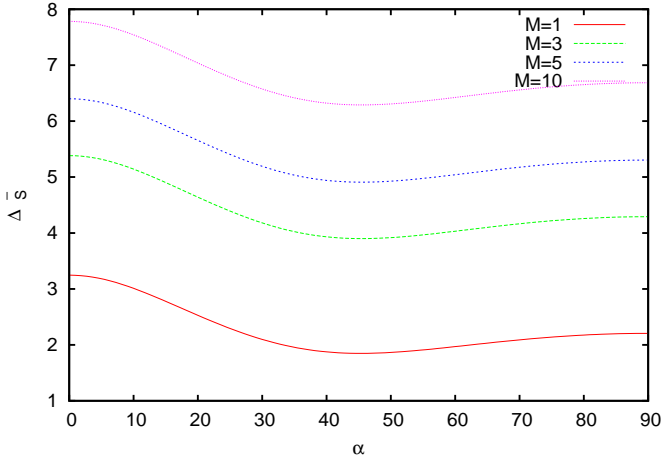}
\caption{The normalized entropy gain as a function of the magnetic field
tilt angle $\alpha$ and the Mach number $M$ for a shock where
electron overshooting is taken into account.
The compression ratio is $s=3$.}
\label{fig-deltas-electrons}
\end{figure}

\begin{figure}
\includegraphics[width=\columnwidth]{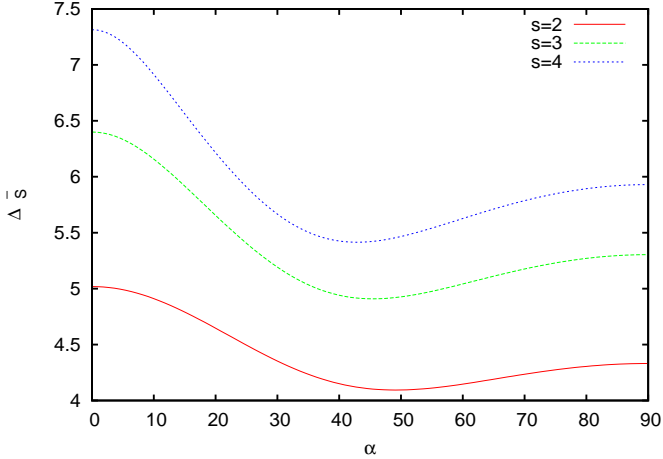}
\caption{The normalized entropy gain as a function of the magnetic field
tilt angle $\alpha$ and the compression ratio $s$ for a shock where
electron overshooting is taken into account.
The mach number is $M=5$.}
\label{fig-deltas-electrons-2}
\end{figure}

Finally, we also want to include the downstream electron pressure
that we have derived in \cite{fs13-multi-p}, joining it with
the ion pressure and derive a more consistent description of the
total particle entropy increase.
In this earlier description, we found that the electron downstream pressure is
substantially increased as a reaction of the negatively charged electrons
to the electric shock potential. Therefore, one can expect that the
electron entropy will be increased as well, and we can start with
Eq. \ref{eq-deltas-magnet-0}, add the electron pressure as found
in the earlier study, and obtain
\begin{equation}
\begin{split}
 \Delta \bar{S}^e
&= \ln [\frac{P_{2,tot}}{P_{1}}(\frac{\rho _{1}}{\rho _{2}})^{\gamma _{adia}}]
\\
&= \ln [\frac{P_{2,p}+P_{2,pui}+P_{2,e}}{P_{1,p}+P_{1,pui}+P_{1,e}}(\frac{\rho _{1}}{\rho _{2}})^{\gamma _{adia}}].
\end{split}
\end{equation}
Now, reminding that $P_{1,p} \simeq P_{1,e}$, we can adopt
$P_{1,p} + P_{1,pui} + P_{1,e} = P_{1,p}(2 + \zeta K)$ and obtain
to the expression
\begin{equation}
\begin{split}
 \Delta \bar{S}^e
=& \ln \left[\frac{s^{-\gamma _{adia}}}{2+\zeta K}(\frac{s}{3}(2A(\alpha)+B(\alpha ))(1+\zeta K) \right.
\\
 &  \left. +\frac{s^{2}-1}{s}\frac{U_{1}^{2}}{c_{1,p}^{2}}\left[\sin ^{2}\alpha A(\alpha )+\cos ^{2}\alpha B(\alpha )\right] \right],
\end{split}
\end{equation}
where the final identity follows from \cite{fs13-multi-p}. Here, $U_1/c_{1,p}$ is the
ratio of the upstream plasma flow speed and the thermal proton sound speed (i.e.
a sonic Mach number).

As one can see in this expression for the total entropy jump related to the particles,
the dependence on the PUI abundance $\zeta$ now reappears, since not all
particles, i.e. electrons and ions, react alike when passing over the shock.
However, as it turns out, the term with the resulting $\zeta$-dependence is negligible
compared to the electron term; in fact, assuming that $\zeta K \gg 1$,
the dependence on these parameters drops out completely.
Different from ions, electrons experience a strong overshooting at the shock,
resulting in a strong heating by thermalisation of this kinetic overshoot energy.
On the other hand, the most important point here is that the magnitude of the entropy
jump now has increased significantly due to the large downstream electron
pressure as shown in Fig. \ref{fig-deltas-electrons}, where we demonstrate that
electrons in fact provide the strongest contribution to the entropy increase,
and that the Mach number $M=U_1/c_{1,p}$ provides the strongest influence on
the overall entropy jump. In addition to this, we have also studies the
dependence on the MHD compression ratio $s$, where one easily sees that the
influence of this parameter is also strong (Fig. \ref{fig-deltas-electrons-2}).

Unfortunately, there is neither data available to check on this
point, nor is there any running
mission dedicated to TS electrons, so it is impossible to verify this result observationally in the
forseeable future. Nevertheless, our results suggest that the electron component
possesses a strong dependence on various parameters of the shock that can be
difficult to observe otherwise, so any future mission to the solar wind TS
would greatly benefit from a dedicated electron instrument.

\subsection{The entropy jump using a downstream $\kappa$ function for a combined proton component}

\label{sec-kappa}

Finally, we study the impact of using a downstream $\kappa$ function to represent
the joint thermal and PUI proton populations. In Sect. \ref{sec-joint}, we
introduced joint downstream proton distribution function describing both
a thermal core and nonthermal tail of PUI protons. Since the parameters of
the $\kappa$-function were set up in a way that the pressure is identical to
the magneto-adiabatic entropy (see Sect. \ref{sec-magnet}), we can not expect new
results from this side.
Instead, we now present some selected, more general aspects of entropy jumps
with $\kappa$ functions at the shock.

First, we want to remind the reader that $\kappa$-functions originate in
a model function for data fits \citep{vasyliunas-68-kappa}, and even after all of the
progress that has been made with understanding $\kappa$-functions,
one still commonly finds modeling approaches where this function
and its parameters are not strongly supported by theoretical arguments.
Therefore, we now close our study of the multicomponent TS with a brief overview
of entropy jumps in $\kappa$-functions.

Taking Eq. \ref{eq-entropy-kappa} and the discussion following it, we can
find the following relation between the Maxwellian and the $\kappa$-entropy:
\begin{equation}
 \bar{S}^{\kappa }
=  S^{0} + \ln \{ \kappa^{3/2} \frac{\Gamma (\kappa-1/2)}{\Gamma (\kappa +1)}\}.
\label{eq-entropy-kappa-2}
\end{equation}
This equation easily proves an important point. As long as the upstream
and downstream $\kappa$-values remain the same, the contribution
to the entropy jump cancels out, and the entropy increase simplifies
to the classical Maxwellian expression $S^0$. Only when there are different
$\kappa$-indices on both sides of the termination shock, one obtains an
additional contribution to the power indices:
\begin{equation}
\begin{split}
\Delta \bar{S}^\kappa &= \Delta \bar{S}^0
 + \ln \left[ \left( \frac{\kappa_2}{\kappa_1} \right)^{3/2}
     \frac{\Gamma(\kappa_2-\frac{1}{2})}{\Gamma(\kappa_1-\frac{1}{2})}
     \frac{\Gamma(\kappa_1+1)}{\Gamma(\kappa_2+1)} \right]
\\
 &= \Delta \bar{S}^0
 + \ln \left[ \sqrt{ \frac{\kappa_2}{\kappa_1} }
     \frac{\Gamma(\kappa_2-\frac{1}{2})}{\Gamma(\kappa_1-\frac{1}{2})}
     \frac{\Gamma(\kappa_1)}{\Gamma(\kappa_2)} \right]
\end{split}
\label{eq-deltas-kappa}
\end{equation}
This relation holds true for arbitrary physical values of $\kappa$ and clearly
demonstrates that the additional contribution to $\Delta \bar{S}$ depends
only on the $\kappa$-parameters on both sides of the shock. In addition,
Eq. \ref{eq-deltas-kappa} is only applicable to systems where the number
of individual shock components does not change between both sides, i.e.
it is not applicable to the joint downstream model adopted in
Sect. \ref{sec-joint}.

\subsection{An eye-guide estimate of the thermodynamically permitted total entropy jump}

The total entropy jump resulting from the full conversion of the free
kinetic energy of the upstream flow can, however, compared to the above
considerations, be estimated much easier along the following procedure: The
normalized maximum entropy jump per particle, $\Delta \bar{S}_{\max }$,
namely is given by the following thermodynamic expression
\begin{equation}
\Delta \bar{S}_{\max }=\frac{Q_{1,2}}{k_B T_{2}}=\frac{\frac{1}{2}%
m(U_{1}^{2}-U_{2}^{2})}{k_{B}T_{2}}\frac{1}{s}=\frac{mU_{1}^{2}(1-\frac{1}{%
s^{2}})}{2sk_{B}T_{2}},
\end{equation}
where $T_{2}$ denotes the effective downstream temperature of the plasma
mixture that absorbs the converted kinetic energy. Taking Voyager-2 shock
crossing data , i.e. $T_{2}=2\cdot 10^{5}K;s=2.5;U_{1}=4\cdot 10^{7}cm/s$,
one then would find
\begin{equation}
\Delta \bar{S}_{\max }=16.
\end{equation}
Comparing this with the values $\Delta \bar{S}_{p}$ and $\Delta \bar{S}%
_{pui} $ which we have displayed in our Figures
\ref{fig-deltas-magnet}, \ref{fig-deltas-wu}, \ref{fig-deltas-electrons} and
\ref{fig-deltas-electrons-2}, one can see that one could easily allow
an increase of the effective downstream temperature by a factor of 80 to 100.
Thus one could allow for instance to increase the
downstream pressure by about this factor without violating any fundamental
thermodynamic law. If, for instance, the downstream thermal ensemble is
characterized by a temperature of the order of $T_{2}=2\cdot 10^{7}K$
(instead as for normal solar wind protons of $2\cdot 10^{5}K$) still
everything would be in complete thermodynamic order without violating
fundamental principles. Hence what we may learn from that view: The strongly
heated downstream electrons, which in our approach \citep{fs13-multi-p}
do attain temperatures of the order of $T_{e,2}\geq 10^{7}K$, are in
excellent agreement with thermodynamically allowed values, and they also
support the maximum entropy principle much better than a pure proton plasma.

If, instead of $T_{2}=2\cdot 10^{5}$~K, one would assume the effective thermal
downstream plasma mixture, due to the strongly heated electrons, to be of
the order of $T_{2}=2\cdot 10^{7}$~K (instead of just the electron temperature $T_{e,2}$),
it then would bring the maximum entropy
jump just down to the achieved level $\Delta \bar{S}_{\max }^{e}=0.16$ (see Figs.
\ref{fig-deltas-magnet} and \ref{fig-deltas-wu}).

\section{Conclusions}

In this article we have investigated how much particle-specific entropy is
produced at the passage of the multifluid solar wind plasma over the MHD
termination shock. Hereby we have started from a consistent solution of the
multifluid MHD Rankine-Hugoniot shock relations to first find the
consistent value of the resulting compression ratio $s$. With the additional
help of kinetic informations on the behaviour of the particle velocity
components at the shock passage and with the use of the Liouville theorem we
then obtain expressions for the downstream distribution functions and the
pressures of the different fluids like solar wind protons, pick-up protons
and electrons. Using then standard thermodynamic expressions we can
calculate fluid-specific entropy productions from the individual fluid
pressures. As we can show then the calculated entropy productions per
particle, both of solar wind protons and pick-up protons, amount to much
lower values than allowed by  thermodynamically maximal values of $\Delta
S_{\max }/k_{b}=16$. Only when including the electron fluid as the
downstream fluid with by far the highest temperature, we can then calculate
for the first time a reasonably high value for the joint entropy production
of the shocked multifluid plasma. Not only do the shocked electrons
represent the most important part of the whole entropy production, they also
different from all earlier representations are shown to be that plasma fluid
with the highest thermal pressure. This fact allows many interesting new
conclusions concerning the dynamics of the downstream heliosheath plasma
flow.

Even though the main outcome of this article here can be seen in the fact
that the decisive part of the entropy production at the plasma passage over
the shock is represented by the strongly heated downstream electrons, we
also want to emphasize the related earlier result from \cite{fs13-multi-p} who
found that these latter, nearly massless particles do also represent the dominant
contribution to the total downstream plasma pressure (and thus, the entropy,
as seen from Eq. \ref{eq-delta-s-mhd}). This eminent feature
has some important consequences for the form, how the downstream plasma
flow organizes itself, as we shall demonstrate below.

As is well known from the equation of motion of the multifluid plasma
mixture, one can construct a typical streamline-constant for the downstream
plasma flow, called the Bernoulli constant $C_{B}$, with the property
$(\vec{U}\cdot \mathbf{grad}C_{B})=0$. In case of the multifluid plasma which
we have considered in this article we find $C_{B}$ as given by
\citep[see][]{bk-landaulifshitz}
\begin{equation}
C_{B}=(1/2)\rho U^{2}+\sum_{i}P_{i}
\end{equation}
where $\rho =\sum_{i}\rho _{i}$ denotes the total mass density of the plasma
and $P_{i}$ denote the different downstream pressures of SW protons, PUI
protons and electrons. Taking now into account that amongst the downstream
pressures the electron pressure $P_{e,2}$ is by far dominant, one can write
for the stagnation streamline, i.e approaching from downstream of the
termination shock the stagnation point of the heliosheath flow at the
heliopause, the following relation
\begin{equation}
C_{B,s}=(1/2)\rho U^{2}+P_{e}=P_{e,s},
\end{equation}
where $P_{e,s}$ is the electron pressure at the stagnation point. This
relation can easily be rearranged to
\begin{equation}
1+\frac{2P_{e}}{\rho U^{2}}=\frac{2P_{e,s}}{\rho U^{2}}.
\end{equation}
Introducing now the effective sound velocity $c_{s}$ and the effective Mach
number in the heliosheath flow by
\begin{equation}
c_{s}^{2}=\frac{\partial P}{\partial \rho }=\frac{\partial }{\partial \rho }%
P_{e}=\frac{\partial }{\partial \rho }(\frac{\rho }{m_{p}}KT_{e})=\frac{P_{e}%
}{\rho }
\end{equation}
and
\begin{equation}
M_{s}=\frac{U}{c_{s}}
\end{equation}
then brings the above relation into the following form
\begin{equation}
1+\frac{2}{M_{s}^{2}}=\frac{2P_{e,s}}{\rho U^{2}}.
\end{equation}
Finally, reminding that $c_{s}=\sqrt{2KT_{e}/m_{p}}=6\ast 10^{7}$~cm/s then shows
that $M_{s}^{2}\approx 10^{-2}$ and that hence in the above relation the
first term on the left side, i.e. $"1"$, can be neglected, then leads to
the simple requirement
\begin{equation}
P_{e}=P_{e,s},
\end{equation}
and with a polytropic relation of gas and density like $C_{p}=P/\rho
^{\gamma }$ then simply states that $\rho /\rho _{s}=1^{-\gamma }=1$, i.e.
that the density along the stagnation streamline, to be generalized to other
streamlines , is constant, and that the plasma consequently behaves
incompressible, which then for instance allows a flow potential $\Phi $ to
be used as \cite{fahr-fichtner-91-helio} have done to describe the heliosheath
streamlines through $\rho \vec{U}=-\mathbf{grad}\Phi $.

\begin{acknowledgements}
M. Siewert is grateful to the Deutsche Forschungsgemeinschaft for financial support granted to him in the frame of the project Si-1550/2-2.
\end{acknowledgements}

\bibliographystyle{aa}
\bibliography{aniso,theory-plasma,experiment-plasma,voyager,ibex,bk-phys,my_papers}

\end{document}